\documentstyle[aps]{revtex}

\newcommand{\be}{\begin{equation}}
\newcommand{\ee}{\end{equation}}
\newcommand{\beq}{\begin{eqnarray}}
\newcommand{\eeq}{\end{eqnarray}}

\newcommand{\eps}{\varepsilon}

\newcommand{\tE}{\lefteqn{\smash{\mathop{\vphantom{<}}\limits^{\;\sim}}}E}
\newcommand{\Et}{\lefteqn{\smash{\mathop{\vphantom{\Bigl(}}\limits_{\sim}
\atop \ }}E}

\newcommand{\tN}{\lefteqn{\mathop{\vphantom{\Bigl(}}\limits_{\sim}
\atop \ }N}
\newcommand{\tM}{\lefteqn{\mathop{\vphantom{\Bigl(}}\limits_{\,\sim}
\atop \ }M}
\newcommand{\tNn}{\lefteqn{\mathop{\vphantom{\Bigl(}}\limits_{\sim}
\atop \ }{\cal N}}

\newcommand{\ng}{{\cal N}_G}
\newcommand{\nl}{{\cal N}_L}
\newcommand{\nd}{{\cal N}_D}

\newcommand{\Fg}{\Phi^G}
\newcommand{\Fl}{\Phi^L}
\newcommand{\Fd}{\Phi^D}
\newcommand{\Fh}{\Phi^H}
\newcommand{\FF}{\Phi}

\newcommand{\E}{\tE^i_a}
\newcommand{\A}{A^a_i}
\newcommand{\ksi}{\xi^a_i}
\newcommand{\ita}{\eta^a_i}

\newcommand{\X}[2]{(\delta_{#1 #2}-\chi_{#1}\chi_{#2})}
\newcommand{\plabel}{\label}
\begin{document}
\draft
\title{Path integral for the Hilbert--Palatini and
Ashtekar gravity}

\author{S.Yu.~Alexandrov\thanks{e.mail: alexand@snoopy.phys.spbu.ru}}

\address{Department of Theoretical Physics,
St.Petersburg University,198904 St.Petersburg, Russia}
\author{D.V.~Vassilevich\thanks{Alexander von Humboldt fellow.
On leave from:
Department of Theoretical Physics,
St.Petersburg University,198904 St.Petersburg, Russia.
e.mail: Dmitri.Vassilevich@itp.uni-leipzig.de}}
\address{Institute for Theoretical Physics, Leipzig
University, Augustusplatz 10/11, 04109 Leipzig, Germany}

\maketitle
\begin{abstract}
To write down a path integral for the Ashtekar gravity
one must solve three fundamental problems.
First, one must understand rules of complex
contour functional integration with holomorphic
action.
Second, one
should find which gauges are compatible with reality
conditions. Third, one should evaluate the Faddeev--Popov
determinant produced by these conditions.
In the present paper
we derive the BRST path integral for the Hilbert--Palatini
gravity. We show, that for certain class of gauge conditions
this path integral can be re-written in terms of the Ashtekar
variables. Reality conditions define contours of integration.
For our class of gauges all ghost terms coincide with
what one could write naively just ignoring any Jacobian
factors arising from the reality conditions.
\end{abstract}
\section{Introduction}
Invention of complex canonical variables \cite{Ash-1} opened
a new avenue for non-perturbative treatment of quantum
general relativity. In these new variables all constraints were
made polynomial at the expense of introducing reality
conditions. Afterwards, many gravitational theories were
re-formulated in a similar way, including even eleven
dimensional supergavity \cite{MeNi}. Quite spectacular
success was achieved in loop quantum gravity \cite{Rov}.
In the view of recent progress of non-perturbative methods
it seems especially important to develop the path integral
formulation of the Ashtekar gravity which could serve as a bridge
between perturbative and non-perturbative results.

Constraint structure of the Ashtekar gravity has been
studied in some detail (for reviews, see \cite{Abook}
and \cite{Peldan}). The BRST charge was constructed
\cite{AshMaTo}. However, this results are still
insufficient for constructing a path integral. It is
known, that any restriction imposed on integration variables
may lead to the Faddeev--Popov ghosts \cite{FaPo}.
It is unclear what kind of ghost action is induced by
the reality conditions.

It is obvious that the path integral for the Ashtekar gravity
will have a somewhat unusual
form. In the case of complex scalar fields action is real and
one integrates over whole complex plane. In the case of Ashtekar
gravity action is holomorphic. Thus one may expect some sort
of contour integration. Position of the contour must be defined
by using the reality conditions. However,
it is not known yet which gauges are compatible with
these conditions.

Our strategy is rather simple. We derive the path integral for
the Hilbert--Palatini gravity and than rewrite it in terms of the
Ashtekar variables. By itself, the first part of our work is
not a great novelty. Hamiltonian structure of the Hilbert--Palatini
gravity has been analyzed in a number of papers
\cite{He,NeTe,ABaJo,Abook,Peldan}.
Given this analysis construction of the path integral is
quite straightforward. However, transition to the Ashtekar variables
requires a complex canonical transformation which is not well
defined in the path integral. We would also like to avoid any
gauge fixing at intermediate steps before the path integral is
written down. Thus we are forced to choose
a basis in the Hilbert--Palatini action different
from the ones used earlier and redo
calculations of the constraint algebra, BRST charge, etc.
A price to pay for the relatively easy transition to the
Ashtekar variables in the path integral is an ugly form
of the Hamiltonian constraint of the Hilbert--Palatini
action. It leads to lengthy calculations at intermediate
steps, which are reported here in some detail to make
the paper self-contained.

As our main result, we transformed the Hilbert--Palatini path integral
to the Ashtekar variables. This can be done successfully
for a restricted class of gauges only. One is not allowed
to impose gauge conditions on the connection variables.
Therefore, path integral quantization of the Ashtekar
gravity in an arbitrary gauge remains an open problem.

The paper is organized as follows. In the next section
some preliminary information on the self dual Hilbert--Palatini
action is collected. We introduce variables which will be convenient
for construction of the path integral, re-derive the Ashtekar
action and give some useful equations. In the third section
we re-consider constraint structure of
the Hilbert--Palatini gravity in terms of our variables.
The fourth section is devoted to the BRST quantization
of the Hilbert--Palatini gravity.
In the section V we establish a relation between first and
second class constraints of the Hilbert--Palatini action
and the reality conditions and vanishing of imaginary part
of the Ashtekar action. In the sixth section we re-write
the path integral in terms of the Ashtekar variables. This
represents our main result. The reader who do not want to go
into technicalities of the BRST quantization will find a
simple derivation of the Faddeev path integral for the Ashtekar
gravity in section VII.
In the last section some
perspectives are briefly discussed. Technical details are
collected in the Appendices.

\section{Selfdual Hilbert--Palatini action}
Let $\Omega^{\gamma\delta}=d\omega^{\gamma\delta}+
{\omega^\gamma}_\alpha \wedge \omega^{\alpha\delta}$, $\omega$
and $e$ are connection and tetrad one-forms respectively.
Signature of the metric is $(-,+,+,+)$. The Levi--Civita
tensor is defined by the equation $\eps_{0123}=1$.
Define the star operator as $\star \omega^{\alpha\beta}=
\frac 12 {\eps^{\alpha\beta}}_{\gamma\delta}
\omega^{\gamma\delta}$. Define
\begin{eqnarray}
A^{\alpha\beta}&=&\frac 12 (\omega^{\alpha\beta}
-i\star \omega^{\alpha\beta}) \plabel{AF} \\
{\cal F}^{\alpha\beta} &=& dA^{\alpha\beta}+{A^\alpha}_\gamma
\wedge A^{\gamma\delta}=\frac 12 (\Omega^{\alpha\beta}
-i\star \Omega^{\alpha\beta}) \nonumber
\end{eqnarray}
These fields satisfy $\star A=iA$, $\star {\cal F} =i{\cal F}$.
Let us start with the selfdual Hilbert--Palatini action expressed
in terms of selfdual connection only \cite{ABaJo,Sam,Ben,Wal}:
\begin{equation}
S_{SD}=\int \eps_{\alpha\beta\gamma\delta}
e^\alpha \wedge e^\beta \wedge {\cal F}^{\gamma\delta}
\plabel{HP1}
\end{equation}
Let us split coordinates $x^\mu$ into "time" $t$ and
"space" $x^i$ and introduce the notations:
\begin{eqnarray}
e^0&=&Ndt+\chi_a E_i^a dx^i ,\quad e^a=E^a_idx^i+E^a_iN^idt
\nonumber \\
A^a_i&=&\eps^{abc}A_{bci} ,\quad
A^a_0=\eps^{abc}A_{bc0} \nonumber \\
F^a_{ij}&=&\eps^{abc}{\cal F}_{ij,bc}
\plabel{split}
\end{eqnarray}
where $a,b,c=1,2,3$ are flat $SO(3)$ indices. $E_a^i$ will denote
inverse of $E_i^a$.
We also need weighted fields:
\begin{equation}
{\tE}^i_a =\sqrt{h}E^i_a,  \quad
\tN=\bigl(\sqrt{h}\bigr)^{-1} N \plabel{wei}
\end{equation}
$\sqrt{h}=\det E^a_i$. After long but elementary calculations we can
represent
 (\ref{HP1}) in the following form
 \begin{eqnarray}
 S_{SD}&=&2 \int
dt\ d^3x (P_a^i\partial_tA_i^a+A_0^a{\cal G}_a
 +N^i{\cal H}_i+\tN{\cal H}) ,
\nonumber \\
 P_a^i&=&i(\tE^i_a-i{\eps_a}^{bc}\tE^i_b\chi_c ) ,
\nonumber \\
{\cal G}_a&=&\nabla_iP^i_a=\partial_iP^i_a
-\eps_{abc}A^b_iP^{ci} , \nonumber \\
{\cal H}_i&=&-2i\tE^k_aF^a_{ik}-\eps_{ijk}\tE^j_a
\tE_b^k\eps^{lmn}\Et^d_l\chi_d F^{ab}_{mn} ,
\nonumber \\
{\cal H}&=&2\tE^i_a\tE^k_b F^{ab}_{ik} ,\plabel{HP2}
\end{eqnarray}
$\Et_i^a=h^{-1/2}E_i^a$.
By a suitable redefinition of Lagrange multipliers $\chi^a$
can be removed from the action.
\begin{equation}
\nd^i=N^i+\frac {\tE^i_a\chi^a(N^j\Et_j^b\chi_b-\tN)}{1-\chi^2}
\qquad {\tNn}=\frac {\tN-N^i\Et_i^a\chi_a}{1-\chi^2}
\plabel{cN}
\end{equation}

The action (\ref{HP2}) now reads:
\begin{eqnarray}
S_{SD}=S_A&=&2\int dt\ d^3x (P_a^i\partial_tA_i^a+A_0^a{\cal G}_a
+{\nd}^i H_i+{\tNn} H) \nonumber \\
 H_i&=&-2P^k_aF^a_{ik}
\nonumber \\
H&=&-2P^i_aP^k_b F^{ab}_{ik} \plabel{HP3}
\end{eqnarray}
All $\chi$-dependence is hidden in the canonical variables.
We arrived at the  Ashtekar action (\ref{HP3}) (later
denoted as $S_A$). Absence of $\chi$ in $S_A$ leads to
a first class primary constraint $p_\chi =0$, where $p_\chi$ is
canonical momentum for $\chi$.
This constraint generates
shifts of $\chi$ by an arbitrary function and originates from the
Lorentz boosts.

One must
bear in mind that not all the components of ${\rm Re}\, P^i_a$
are independent. To restore correct form of $P^i_a$ one needs
a condition ${\rm Im}\, P^{(i}_a {\rm Re}\, P^{j)}_a=0$ or,
equivalently,
\begin{equation}
{\rm Im}\, (P_a^i P_a^j)=0 \label{1rc}
\end{equation}
The equation (\ref{1rc}) is known as first metric reality
condition. Being supplemented by the second metric reality
condition
\begin{equation}
\partial_t {\rm Im}\, (P_a^i P_a^j)=0 \label{2rc}
\end{equation}
on an initial hypersurface it ensures real evolution
of the metric \cite{ARoTa,Imm,YoSh}. As usual, the triad
field $\tE$ should be non-degenerate.

Define the smeared constraints:
\begin{eqnarray}
&&{\cal G}(n)=\int d^3x\, n^a{\cal G}_a, \qquad
H^A(\tN )=\int d^3x\, \tN H
\nonumber \\
&&{\cal D}(\vec N)=\int d^3x\, N^i(H_i+2A_i^a{\cal G}_a),
\end{eqnarray}
They obey the following algebra:
\begin{eqnarray}
&&\left\{ {\cal G}(n) ,{\cal G}(m) \right\}_C=-{\cal G}(n\times m),
\nonumber \\
&&\left\{ {\cal D}(\vec N) ,{\cal D}(\vec M) \right\}_C=
-2{\cal D}([\vec N ,\vec M ]),\nonumber \\
&&\left\{ {\cal D}(\vec N) ,{\cal G}(n) \right\}_C=-
2{\cal G}( N^i\partial_in), \nonumber \\
&&\Bigl\{ H^A(\tN ) ,{\cal G}(n) \Bigr\}_C =0, \plabel{algA} \\
&&\Bigl\{ {\cal D}(\vec N) ,H^A(\tN ) \Bigr\}_C=
-2H^A({\cal L}_{\vec N}\tN ), \nonumber \\
&&\Bigl\{ H^A(\tN ),H^A(\tM ) \Bigr\}_C =
2{\cal D}(\vec K)-2{\cal G}(2K^jA_j) \nonumber
 \end{eqnarray}
where
\begin{eqnarray}
&&(n\times m)^a=\eps^{abc}n^bm^c,\qquad
{\cal L}_{\vec N}\tN =
N^i\partial_i \tN-\tN\partial_iN^i, \nonumber \\
&&[\vec N ,\vec M ]^i=
N^k\partial_kM^i-M^k\partial_kN_i, \plabel{not1} \\
&& K^j=(\tN\partial_i\tM-\tM\partial_i\tN)P^i_aP^j_a
\plabel{not11}
\end{eqnarray}
We introduced the subscript $C$ to distinguish
the Poisson bracket
$\{ \cdot ,\cdot \}_C$
of the complex Ashtekar theory from
that of the real Hilbert-Palatini action.

\section{Hamiltonian form of the Hilbert--Palatini action}

Let us start with the Hilbert--Palatini action
\be S=\frac 12\int \eps_{\alpha\beta\gamma\delta}
 e^\alpha \wedge
e^\beta \wedge {\Omega}^{\gamma\delta} \ee
Recall that the
Ashtekar action is obtained from the Hilbert--Palatini
 one by adding a pure imaginary term
$-i\frac 12 \int \eps_{\alpha\beta\gamma\delta}
 e^\alpha \wedge e^\beta
\wedge \star {\Omega}^{\gamma\delta}$.
Therefore,
\be S={\rm Re} \ S_{A}=2\int dt\ d^3x (\tE_a^i\partial_t
\omega_i^{0a}+Z^i_a\partial_t \xi_i^a +
n_G^a{\rm Re}\,{\cal G}_a+n_L^a{\rm Im}\,{\cal G}_a
 +{\nd}^i
{\rm Re}\,H_i+{\tNn} {\rm Re}\,H) \ee
where \beq && n_G^a={\rm Re}\, A_0^a,\quad
n_L^a=-{\rm Im}\, A_0^a \\
&&Z_a^i={\eps_a}^{bc}\tE^i_b \chi_c\\
&&\xi_i^a= \frac 12 {\eps^a}_{bc}\omega_i^{bc}
 \eeq

In order to simplify the constraint algebra
we replace ${\rm Re}\, H_i$ by the modified vector constraint.
To this end
we shift the Lagrange multipliers.
 \be   n_G^a = \ng^a + 2\nd^i\xi_i^a \ \ \ \
n_L^a = \nl^a+2\nd^i\omega_i^{0a}   \ee

We see that $\E$ plays a role of the momentum for $\xi_i^a$ whereas
$Z^i_a$ is momentum conjugate to
 $\omega_i^{0a}$.
$Z_a^i$ has three independent components only. To have time derivatives
of true dynamical variables we replace
\be \omega_i^{0a} = \eta^a_i +\eps^{abc}\xi_i^b \chi_c \ee
Then the kinetic term reads $\E \partial_t \ita -
(\eps^{abc}\xi^b_i \tE^i_c) \partial_t \chi_a$. By a suitable
change of variables we can bring this term to the standard
form $p\partial_t q$.
Let us introduce a basis in the space of $3\times 3$ matrices.
\be (r_A)_i^a= \Et_i^b (\beta_A)^a_b, \quad
(\gamma_a)_i^b=\frac 12 \eps_{abc}\Et_i^c \plabel{bas}  \ee
where $\beta_A$ are six symmetric $3\times 3$ matrices.
Define
\be  \ksi = r_i^a +(\gamma_b)^a_i \omega^b, \qquad
r_i^a=(r_A)^a_i \lambda^A  \label{ksi} \ee
$\omega$ and $\lambda$ will be treated as new canonical
variables.

We arrive at the following expression for the Hilbert--Palatini
action
\begin{eqnarray}  \frac 12 S &=&\int dt\ d^3x (\E\partial_t \ita+
\chi_a\partial_t \omega^a + \ng^a \Fg_a+\nl^a\Fl_a +\nd^i \Fd_i+\tNn \Fh)
\plabel{HP5} \\
\Fg_a &=& \partial_i ({\eps_a}^{bc} \tE^i_b \chi_c)-{\eps_{ab}}^c \eta_i^b
\tE^i_c -{\eps_{ab}}^c \omega^b \chi_c \nonumber \\
\Fl_a &=& \partial_i\E +\eps_{abc}\eta_i^b \eps^{cgf}\tE^i_g \chi_f -
(\delta_{ab}-\chi_a \chi_b)\omega^b \nonumber \\
\Fd_i &=&-2\left[ \tE^j_a \partial_i \eta_j^a -\partial_j(\tE^j_a\eta_i^a)
- \omega^a \partial_i \chi_a \right] \nonumber \\
\Fh &=&  \eps^{abc}\tE_b^i
\tE_c^j (\delta_{ad}-\chi_a \chi_d) {\eps^d}_{gf} \eta_i^g \eta_j^f +
2\tE^i_a \tE^j_b \chi^b (\partial_i \eta_j^a-\partial_j
\eta_i^a) \nonumber \\
&& -(1-\chi^2)(2\partial_i (\tE^i_a\omega^a)
-h^{-1}\omega^a \partial_i (h\tE^i_a))+\omega^a
\chi^b(\tE^i_a\partial_i\
\chi^b +\tE_b^i\partial_i \chi_a) \nonumber \\
&&+ \tE^j_a\omega^b(\chi_a\eta_j^b-\chi_b \eta_j^a)- \omega^a
\chi_a(\tE^j_b\eta_j^c\chi^b\chi_c - \chi^2\tE_b^j \eta_j^b)  \nonumber \\
&&-\frac 12 (1-\chi^2)\omega^a \omega^b (\delta_{ab}-\chi_a \chi_b)
\nonumber \\
&&+ 2\eps^{abc}\tE^i_b \tE^j_c \left( (1-\chi^2)\partial_i r_j^a+
r_j^d \chi_d \partial_i \chi_a -(1-\chi^2)\chi_a r_j^d\eta_i^d
\right. \nonumber \\
&&\left. +(\delta_{ag}-\chi_a \chi_g)\eta_i^g r_j^d \chi_d \right) -
(1-\chi^2)\eps^{abc} \tE^i_b\tE_c^j (\delta_{ad}-\chi_a \chi_d)
{\eps^d}_{gf} r_i^g r_j^f \nonumber
\end{eqnarray}

We see that $\lambda_A$ has no conjugate
momentum, and thus  is non-dynamical. We observe also that $\lambda_A$
is contained in $\Phi^H$ only.

Let us analyse constraints of the theory along the lines of usual
Dirac procedure \cite{Dirac}.
Since all steps are completely standard we omit irrelevant
technical details (cf. \cite{ABaJo,Abook}).
First we note that
$\E$ and $\chi_a$ are conjugate momenta to $\ita$ and
$\omega^a$ respectively. By analyzing the consistency conditions we
get the following set of constraints
\be p^{(n)}_{\alpha }=0 \qquad p^{(\lambda )}_A=0 \qquad
\FF_{\alpha }=0  \qquad \tNn \frac {\partial \Fh}{\partial \lambda_A}=0 \ee
where $p^{(q)}$ denotes momentum conjugate to variable $q$,
$(n)$ are all Lagrange multipliers, and $\FF_{\alpha}=
(\Fg_a,\Fl_a,\Fd_i,\Fh)$.
Introduce
\begin{equation}
 \FF_{\alpha}^{'}=\FF_{\alpha}-\frac 12 p^{\lambda}_A
{\cal A}_{AB}^{-1} \left\{ \FF_{\alpha},\frac{\partial \Fh}{\partial
\lambda_B}\right\} ,\end{equation}
where ${\cal A}_{AB}=-\frac 12
\frac {\partial^2 \Fh} {\partial \lambda_A \partial
\lambda_B}$. Then  $\FF^{'}_{\alpha}$ and
$p^{(n)}_{\alpha}$ are first class constraints.

Remaining constraints $p_A^{(\lambda )}$ and
$\tNn \frac {\partial \Fh}{\partial \lambda_A}$
are second class constraints with
nontrivial matrix of commutators.
This matrix
is non-degenerate and can be used to construct Dirac's bracket. To avoid
using such an object one should solve second class constraints explicitly.

The constraints $p^{(\lambda )}_A=0$ are solved trivially giving
us back $\Phi_\alpha$ as first class constraints.
Since $\Phi^H$ is quadratic in $\lambda$, it can be represented
as
\begin{equation}
 \Fh =\Fh_0+2{\cal B}_A\lambda_A
-\lambda_A {\cal A}_{AB}\lambda_B,
\end{equation}
The remaining second class constraints give the equations
\begin{equation}
0=\frac {\delta \Phi^H}{\delta\lambda^A} =
2(-{\cal A}_{AB}\lambda_B+{\cal B}_A) ,\plabel{2class}
\end{equation}
which can be solved for $\lambda$ resulting in
expressions for non-dynamical components $r_i^a$  in terms of other
canonical variables.
Here we give final results only.
Some intermediate steps are reported in the Appendix A.
\begin{eqnarray}
r_i^a&=&\frac 1{2(1-\chi^2)} \Bigl( -X_{ad} \eps^{dbc}
\tE_b^k \tE^j_c X_{gf} \Et_i^g \partial_k \Et_j^f \Bigr.
\nonumber \\
&&+X_{ag} \Et_i^g \eps^{dbc} \tE^k_b \tE^j_c X_{df} \partial_k\Et_j^f
-\eps^{dbc} \tE_b^k\tE_c^j X_{dg} \Et_i^g X_{af}\partial_k\Et_j^f
\nonumber \\
&&-\chi_a \eps_{dbc} \tE^j_b X_{cg} \Et_i^g \partial_i \chi_d
+\eps^{abc}\chi_b\partial_i\chi_c
-\eps^{abc}\chi_b \tE_c^j\Et_i^d\partial_j\chi_d \nonumber \\
&&\Bigl. +\eps^{abc}\chi_b \eta_i^c
+\eps^{dbc}\tE_a^j\Et_i^d \chi_b \eta_j^c \Bigr) ,\label{rai}
\end{eqnarray}
where $X_{ab}=(\delta_{ab}-\chi_a\chi_b )$.
The Hamiltonian constraint reads:
\begin{eqnarray}
\Fh&=&\Fh_0+ {\cal B}_A {\cal A}_{AB}^{-1} {\cal B}_B
\nonumber \\
&=&-\frac12(1-\chi^2)\omega^a\omega^b X_{ab}
 -(1-\chi^2)\left( 2\partial_i(\tE^i_a\omega^a)-\right. \nonumber \\
&&h^{-1}\omega^a\partial_i(h\tE^i_a))
+\omega^a\chi_b(\tE^i_a\partial_i\chi_b
+\tE^i_b\partial_i\chi_a) \nonumber \\
&&+\left. (\tE^i_a\omega^b(\chi_a\eta_i^b-\chi_b\eta_i^a)-\omega^a\chi_a
(\tE^i_b\chi^b\eta_i^c\chi_c-\chi^2\tE^i_b\eta_i^b)\right) \nonumber \\
&&+\frac12\Bigl\{ -\eps^{abc}\tE^i_b\tE^j_c X_{ad}\eps^{dpq}\tE^k_p\tE^l_q
X_{gf}\partial_i\Et_j^g\partial_k\Et_l^f+\Bigr.\nonumber \\
&&\eps^{abc}\tE^i_b\tE^j_c X_{ag}\partial_i\Et_j^g
\eps^{dpq}\tE^k_p\tE^l_q X_{df}\partial_k\Et_l^f  \nonumber \\
&&\Bigl. -\eps^{abc}\tE^i_b\tE^j_c X_{ag}\partial_k\Et_l^g
\eps^{dpq}\tE^k_p\tE^l_q X_{df}\partial_i\Et_j^f  \Bigr\}\nonumber \\
&&+\Bigl\{ -\eps^{abc}\tE^i_b\tE^j_c \chi_a\eps^{dpq}\tE^k_p\partial_k\chi_d
X_{qg}\partial_i\Et_j^g+ \eps^{abc}\tE^i_b\tE^j_c \partial_i\Et_j^a
\eps^{dpq}\tE^k_p \partial_k\Et_l^f \chi_q \Bigr. \nonumber \\
&&\Bigl. - \eps^{abc}\tE^i_b\tE^j_c \eps^{adp}\tE^k_p\partial_k\chi_d
\partial_i\Et_j^g\chi_g  \Bigr\}\nonumber \\
&&-\frac{\chi^2}{2(1-\chi^2)}\eps^{abc}\tE^i_b\partial_i\chi_a X_{cq}
\eps^{dpq}\tE^j_p\partial_j\chi_d \nonumber \\
&&+\eps^{abc}\tE^k_a\tE^i_b\tE^j_c\eps^{dpq}\chi_d\eta_k^p\partial_i\Et_j^q
-\eps^{abc}\tE^i_b\partial_i\tE^j_c\eps^{apq}\chi_p\eta_j^q \nonumber \\
&&+\frac{1}{1-\chi^2}\eps^{abc}\tE^i_b\partial_i\chi_a
\eps^{cpq}\chi_p\eta_j^q\tE^j_d\chi_d \nonumber \\
&&+\Bigl\{ 2\tE^i_a\tE^j_b\chi_b(\partial_i\eta_j^a-\partial_j\eta_i^a)+
\eps^{abc}\tE^i_b\tE^j_c\eps_{apq}\eta_i^p\eta_j^q+\frac12\eps^{abc}
\chi_a\tE^i_b\eta_j^c\eps^{dpq}\chi_d\tE^j_p\eta_i^q\Bigr. \nonumber \\
&&-\Bigl.\frac12\eps^{abc}\chi_a\eta_i^b
\eps^{cpq}\chi_p\eta_j^q\tE^i_g\tE^j_g-
 \frac1{2(1-\chi^2)}\eps^{abc}\chi_a\eta_i^b\eps^{cpq}\chi_p\eta_j^q
\tE^i_g\chi_g\tE^j_f\chi_f \Bigr\}  . \plabel{newH}
\end{eqnarray}

We end up this section with some useful commutators. Introduce
smeared first class constraints:
\begin{eqnarray}
&&G(n)=\int d^3x\, n^a\Fg_a, \quad L(m)=\int d^3x\, m^b \Fl_b,
\nonumber \\
&&D(\vec N)=\int d^3x\, N^i\Fd_i, \quad H(\tN )=\int d^3x\, \tN \Fh
\end{eqnarray}
Here all the constraints are taken from (\ref{HP5}), except for
the Hamiltonian constraint $\Phi^H$ which is now given by
(\ref{newH}).  $\ksi$ is expressed in terms of canonical variables by
means of (\ref{ksi}) and (\ref{rai}).

The transformations of the connection fields are:
\begin{eqnarray}
&&\Bigl\{ G(n) ,\xi_j^d \Bigr\}=\eps^{dab}n^a\xi_j^b+\partial_jn^d,
\nonumber \\
&&\Bigl\{ G(n) ,\eta_j^d+\eps^{dpq}\xi_j^p\chi_q \Bigr\}=
\eps^{dab}n^a(\eta_j^b+\eps^{bpq}\xi_j^p\chi_q),\nonumber \\
&&\Bigl\{ L(m) ,\xi_j^d \Bigr\}=
-\eps^{dab}m^a(\eta_j^b+\eps^{bpq}\xi_j^p\chi_q), \nonumber \\
&&\Bigl\{ L(m) ,\eta_j^d+\eps^{dpq}\xi_j^p\chi_q  \Bigr\}=
\eps^{dab}m^a\xi_j^b+\partial_jm^d,\nonumber \\
&&\Bigl\{ D(\vec N) ,\xi_j^d \Bigr\}=
2(N^i\partial_i\xi_j^d+\xi_i^d\partial_jN^i),\plabel{comm} \\
&&\Bigl\{ D(\vec N) ,\eta_j^d+\eps^{dpq}\xi_j^p\chi_q  \Bigr\}=
2(N^i\partial_i(\eta_j^d+\eps^{dpq}\xi_j^p\chi_q)
+(\eta_i^d+\eps^{dpq}\xi_i^p\chi_q)\partial_jN^i)
\nonumber
\end{eqnarray}
The Poisson brackets between the constraints are straightforward
to evaluate. One obtains
\begin{eqnarray}
&&\Bigl\{ G(n) ,G(m) \Bigr\}=-G(n\times m), \nonumber \\
&&\Bigl\{ L(n) ,L(m) \Bigr\}=G(n \times m) , \nonumber \\
&&\Bigl\{ G(n) ,L(m) \Bigr\}=-L(n\times m) ,  \nonumber \\
&&\left\{ D(\vec N) ,D(\vec M) \right\}=-2D([\vec N ,\vec M ]),
 \nonumber \\
&&\left\{ D(\vec N) ,G(n) \right\}=-2G( N^i\partial_in),
 \nonumber \\
&&\left\{ D(\vec N) ,L(m) \right\}=-2L(N^i\partial_im),\plabel{alg}\\
&&\Bigl\{ H(\tN ) ,G(n) \Bigr\} =0,  \nonumber \\
&&\Bigl\{ H(\tN ) ,L(m) \Bigr\} =0,  \nonumber \\
&&\Bigl\{ D(\vec N) ,H(\tN ) \Bigr\} =-2H({\cal L}_{\vec N}\tN ),
 \nonumber \\
&& \Bigl\{ H(\tN ),H(\tM ) \Bigr\} = 2D(\vec K)-2G(2K^j\xi_j)-
 2L(2K^j(\eta_j+\xi_j\times \chi)) \nonumber
\end{eqnarray}
where
\begin{eqnarray}
&& K^j[\tN ,\tM]= (\tN \partial_i\tM -\tM \partial_i\tN )K^{ij}
\nonumber \\
&&K^{ij}=-
(\tE^i_a\tE^j_a(1-\chi^2)+\tE^i_a\chi_a\tE^j_b\chi_b). \plabel{Kij}
\end{eqnarray}
Other notations are taken from (\ref{not1}). $K^{i}$ is in fact
the same as in (\ref{not11}) but written in different variables.

$\Phi^H$ will be called the Hamiltonian constraint. $\Phi^D$
generates diffeomorphisms of the 3-surface and will be called
the diffeomorphism constraint. $\Phi^G$ and $\Phi^L$ generate
the $SO(3,R)$ rotations and the Lorentz boosts respectively.
They will be called the Gauss law constraint and the Lorentz
constraint, respectively.

There is a set of remarkable relations between the Poisson
brackets of the Hilbert--Palatini gravity and that of the
Ashtekar gravity.
\begin{eqnarray}
&&\{ {\cal G}(n),P_a^j\}_C=\{ G(n),P_a^j\}=
\{ iL(n),P_a^j\} , \nonumber \\
&&\{ {\cal G}(n),A_j^a\}_C=\{ G(n),A_j^a\}=
\{ iL(n),A_j^a\} , \nonumber \\
&&\{ {\cal D}(\vec N), P_a^j\}_C=\{ D(\vec N),P_a^j\} ,
\quad
\{ {\cal D}(\vec N), A^a_j\}_C=\{ D(\vec N),A^a_j\} ,
\nonumber \\
&&\{ H^A(N),P_a^j\}_C=\{ H(N),P_a^j\} \plabel{prop}
\end{eqnarray}
Note, that last relation holds for $P_a^j$ only.

In a different context relation between Hilbert--Palatini
and Ashtekar brackets was considered recently by
Khatsymovsky \cite{Kh}.

\section{BRST quantization of the Hilbert--Palatini gravity}

 In this section we construct the BRST path integral \cite{BFV}
for the Hilbert--Palatini gravity. Here we follow the review
 \cite{Henneaux}. Consider a dynamical system with  phase space
variables $(q^s,p_s)$,
Hamiltonian $H_0$, and constraints $\FF_{\alpha}$.
Let $n^{\alpha}$ be the Lagrange multipliers associated with
the constraints $\FF_{\alpha}$, and $\pi_{\alpha}$ be the canonically
conjugate momenta. The extended phase space is defined by introducing
extra ghost and antighost fields
$(b^{\alpha},\bar c_{\alpha},c^{\alpha},\bar b_{\alpha})$.
obeying the following nonvanishing
antibrackets
$$ \{ b^{\alpha} ,\bar c_{\beta} \}_+=-\delta^{\alpha}_{\beta},\
\{ c^{\alpha} ,\bar b_{\beta} \}_+=-\delta^{\alpha}_{\beta} $$
 $c^{\alpha},\bar c_{\alpha}$ are real, whereas $b^{\alpha},\bar b_{\alpha}$
are imaginary.

It is convenient to define an additional structure on the extended
phase space, that of "ghost number". This is done by attributing
the following ghost number to the canonical variables:
 $c^{\alpha},b^{\alpha}$ have ghost number one,
 $\bar c_{\alpha},\bar b_{\alpha}$ have ghost number minus one.
All other variables have ghost number zero.

On this space one can construct a BRST generator $\Omega$
and a BRST invariant Hamiltonian $H$. They are determined by the following
conditions:

(a) $\Omega$ is real and odd; (b) $\Omega$ has ghost number one;
(c) $\Omega =-ib^{\alpha}\pi_{\alpha}+c^{\alpha}\FF_{\alpha}+
"higher\  ghost\ terms"$; (d) $\{ \Omega ,\Omega \}_+=0$

(a) $H$ is real and even; (b) $H$ has ghost number zero;
(c) $H$ coincides with $H_0$ up to higher ghost terms;
(d) $\{ H ,\Omega \}=0$

 If $H_0$ weakly vanishes (as in our case) one can take $H=0$ since
the formalism supports an arbitrariness in the definition of observables:
$H_0 \sim H_0+k^{\alpha}\FF_{\alpha}$.

 The BRST generator is fully defined by structure functions of
the constraint algebra:
$$\Omega=-ib^{\alpha}\pi_{\alpha}+\sum\limits_{n\ge 0}c^{\alpha_{n+1}}
\cdots c^{\alpha_1}U^{(n)\beta_1 \cdots \beta_n}_{\alpha_1
\cdots \alpha_{n+1}}\bar b_{\beta_n}\cdots \bar b_{\beta_1}$$

The structure functions for
the Hilbert--Palatini gravity are constructed in the Appendix B.
As a result, we obtain
\be \Omega=-ib^{\alpha}\pi_{\alpha}+c^{\alpha}\FF_{\alpha}+\frac12
c^{\alpha}c^{\beta}C^{\gamma}_{\alpha \beta}\bar b_{\gamma}+
c^{\alpha}c^{\beta}c^{\gamma}U^{(2)\delta \lambda}_{\alpha \beta \gamma}
\bar b_{\delta}\bar b_{\lambda}   \label{BRSg} \ee
where $U^{(2)}$ is taken from (\ref{sf2}). Note that for the Yang--Mills
theory the term with $U^{(2)}$ is absent in the BRST charge. This is
also the case of the Ashtekar gravity \cite{AshMaTo}.

The quantization is based on the generating functional for the Green
functions which is represented in the form
\be Z[j,J,\lambda ] =\int {\cal D}\mu e^{i\int dt\,
(L_{eff}+j_sq^s+J^sp_s+\lambda_{\alpha} n^{\alpha})} \ee
where
\be  L_{eff}=\dot q^s p_s +\dot n^{\alpha}\pi_{\alpha}+
\dot c^{\alpha}\bar b_{\alpha}+\dot b^{\alpha}\bar c_{\alpha}-H_{eff}
\qquad
H_{eff}=H-\{ \psi ,\Omega \}_+ \plabel{Leff} \ee
Here $\psi$ is an odd and imaginary function which has
ghost number minus one and plays a role of  gauge fixing function,
whereas ${\cal D} \mu$ is the usual measure (product over time of
the Liouville measure of the extended phase space).

 Let us choose
\begin{equation}
\psi= -\bar b_{\alpha}n^{\alpha}+i\bar c_{\alpha}
 \bigl( \frac1{\gamma}f^{\alpha}(q,p)+\frac1{\gamma}g^{\alpha}(n)\bigr) .
\plabel{psi}
\end{equation}
By substituting (\ref{BRSg}) and (\ref{psi}) in (\ref{Leff}) and putting
$H=0$ one obtains:
 \beq H_{eff}&=&- n^{\alpha}\FF_{\alpha}-i\bar b_{\alpha} b^{\alpha}+
 c^{\alpha}n^{\beta}C^{\gamma}_{\alpha \beta}\bar b_{\gamma} -
 3c^{\alpha}c^{\beta}n^{\gamma}U^{(2)\delta \lambda}_{\alpha \beta \gamma}
 \bar b_{\delta}\bar b_{\lambda} \nonumber \\
 && +\frac1{\gamma}\Bigl\{ (f^{\alpha}+g^{\alpha})\pi_{\alpha}-
 \bar c_{\alpha}\frac{\partial g^{\alpha}}{\partial n^{\beta}}b^{\beta}-
 i\bar c_{\alpha}\{ f^{\alpha}, \FF_{\beta}\} c^{\beta}-
 i\bar c_{\alpha}\{ f^{\alpha},C^{\delta}_{\beta \gamma}\}
 c^{\beta}c^{\gamma}\bar b_{\delta} \Bigr. \nonumber \\
 && \Bigl. -i\bar c_{\alpha}\{ f^{\alpha},
 U^{(2)\xi \eta}_{\beta \gamma \delta}\} c^{\beta}c^{\gamma}c^{\delta}
 \bar b_{\xi}\bar b_{\eta}    \Bigr\} \eeq

    Let us make the change of variables with unit Jacobian:
    $$ \pi_{\alpha} \longrightarrow \gamma \pi_{\alpha},\ \
       \bar c_{\alpha} \longrightarrow \gamma \bar c_{\alpha} $$
 Then let $\gamma \longrightarrow 0$. In this limit
integration over $\pi_{\alpha},\ b^{\alpha}$ and
 $\bar b_{\alpha}$ is easily performed giving:
\be Z[j,J,\lambda ] =\int {\cal D}q {\cal D}p {\cal D}n {\cal D}c
{\cal D}\bar c \delta (f^{\alpha} {+} g^{\alpha})  e^{i\int dt\,
(L_{eff}'+j_sq^s+J^sp_s+\lambda_{\alpha} n^{\alpha})}  \label{ZHP} \ee
where
\beq L_{eff}'&=&  \dot q^s p_s+n^{\alpha}\FF_{\alpha}-
i\bar c_{\beta}\Bigl( \frac{\partial g^{\beta}}{\partial n^{\alpha}}
\partial_t - \frac{\partial g^{\beta}}{\partial n^{\gamma}}
C^{\gamma}_{\alpha \lambda}n^{\lambda}+\{ \FF_{\alpha} ,f^{\beta}\} \Bigr)
c^{\alpha} \nonumber \\
&& -\bar c_{\xi}\bar c_{\eta}\Bigl( \frac{\partial g^{\eta}}
{\partial n^{\delta}}\{ f^{\xi}, C^{\delta}_{\alpha \beta}\} +
3\frac{\partial g^{\xi}}{\partial n^{\delta}}
\frac{\partial g^{\eta}}{\partial n^{\lambda}}
U^{(2)\delta \lambda}_{\alpha \beta \gamma}n^{\gamma}\Bigr) c^{\alpha}
c^{\beta} \nonumber \\
&& -i\bar c_{\alpha}\bar c_{\xi}\bar c_{\eta}
 \frac{\partial g^{\xi}}{\partial n^{\lambda}}
\frac{\partial g^{\eta}}{\partial n^{\sigma}}\{ f^{\alpha},
U^{(2) \lambda \sigma}_{\beta \gamma \delta}\} c^{\beta}c^{\gamma}c^{\delta}
\plabel{Lsht}
\eeq
and $q^s=(\ita,\omega^a),\ p_s=(\E,\chi_a)$.

This completes construction of the path integral for the Hilbert--Palatini
gravity.One can see
that  dependence of structure constants on canonical variables leads to
appearance of multighost interaction terms in (\ref{Lsht}).
By an appropriate choice of  gauge fixing functions one can eliminate
these terms. All nonvanishing components of $U^{(2)}$ have
upper indices corresponding to the Gauss or Lorentz constraints.
Therefore, if the functions $g^\alpha$ do not depend on the Lagrange
multipliers ${\cal N}_G$ and ${\cal N}_L$, all terms with $U^{(2)}$
disappear. If, furthermore, the functions $f^\alpha$ do not depend
on canonical coordinates $q^s$, the Poisson bracket
$\{ f^{\xi}, C^{\delta}_{\alpha \beta}\}$ vanishes and the remaining
higher ghost terms disappear also. In such a case, general structure
of the path integral is identical to that of rank one Yang--Mills
theory. For short, these gauges will be called the Yang--Mills
(YM) gauges. They play an important role in path integral quantization
of the Ashtekar gravity.

\section{Constraints versus reality conditions}
In this section we establish relation between solutions of
the constraints in the real Hilbert--Palatini formulation
and the reality conditions  (\ref{1rc}) and (\ref{2rc})
of the Ashtekar gravity.
Let us recall expressions for the complex canonical variables
$P$ and $A$ in terms of the real canonical variables:
\begin{eqnarray}
 P_a^i&=&i(\tE^i_a-i{\eps_a}^{bc}\tE^i_b\chi_c ) ,\nonumber \\
A_j^a&=&\xi_j^a -i(\eta_j^a+\eps^{abc}\xi_j^b\chi_c) ,
\nonumber \\
\xi_j^a&=&r_j^a-\frac 12 \eps^{abc}\omega_b \Et_j^c ,
\plabel{CtoR}
\end{eqnarray}
$r_j^a$ is given by the equation (\ref{rai}).

Here it will be demonstrated the reality conditions (\ref{1rc})
and (\ref{2rc}) are satisfied by (\ref{CtoR}) provided the
canonical variables of the real theory satisfy the Gauss law
and the Lorentz constraint. Moreover, we shall prove that the
Ashtekar action is real under the same conditions.
The last statement is not completely trivial even though
real Hilbert--Palatini action is related to complex Ashtekar
action by a canonical transformation. The point is that this
transformation is not canonical on the whole phase space
\cite{Abook}.
Thus for our basis in the phase space reality of the Ashtekar
action must be checked independently.

The first reality condition (\ref{1rc}) is satisfied trivially.
Let us rewrite (\ref{2rc}) in a more explicit form. Time
evolution $P_a^lP_a^j$ is given by Poisson bracket of total
complex Hamiltonian (\ref{HP3}) and $P_a^lP_a^j$:
\begin{eqnarray}
\partial_t\, (P_a^lP_a^j)&=&\left\{
\int dt\ d^3x (A_0^a{\cal G}_a
+{\nd}^i H_i+{\tNn} H),
P_a^lP_a^j \right\}_C \nonumber \\
&=&-2(2P^l_aP^j_a\partial_i{\nd}^i -P^k_aP^l_a\partial_k{\nd}^j
-P^k_aP^j_a\partial_k{\nd}^l +{\nd}^i\partial_i(P^l_aP^j_a))
\nonumber \\
&&+2(\nabla_k P^k_a)({\nd}^jP^l_a+{\nd}^lP^j_a) \nonumber \\
&&-2{\tNn} \eps^{abc}P_a^i (P^j_c\nabla_i P^l_b +
P^l_c\nabla_i P^j_b ) .\plabel{2rc2}
\end{eqnarray}
First line of (\ref{2rc2}) is real for real ${\nd}^i$ due to
the first reality condition (\ref{1rc}). Second line disappears
due to the Gauss law constraint. Therefore, to ensure real
metric evolution one must require
\begin{equation}
{\rm Im}\, (\eps^{abc}P_a^i (P^j_c\nabla_i P^l_b +
P^l_c\nabla_i P^j_b ) )=0 .\plabel{2rc3}
\end{equation}
The condition (\ref{2rc3}) can be presented as ${\rm Im}\,
\{ P_a^lP_a^j,H\}_C =0$. It is clear that this condition
is invariant under {\em complex} $SO(3)$ transformations.
These transformations can be used to put $\chi =0$.
One can easily demonstrate that for the fields (\ref{CtoR})
the condition (\ref{2rc3}) is satisfied.

Now let us prove that under the same conditions
\begin{equation}
{\rm Im}\, H_i = {\rm Im}\, (H_i+2A_i^a{\cal G}_a)=0 .
\plabel{ImHi}
\end{equation}
From the equations (\ref{algA}) and (\ref{prop}) one can see
that $\{ {\cal G},{\cal G}\}_C\sim {\cal G}$ and
$\{ \Phi^D,{\cal G}\}\sim {\cal G}$. Hence the surface ${\cal G}=0$
is invariant under complex $SO(3)$ transformations and real
diffeomorphisms. Since
$\{ {\cal G}, H_i+2A_i^a{\cal G}_a\}_C \sim {\cal G}$
and $\{ \Phi^D, {\rm Im}\, (H_i+2A_i^a{\cal G}_a) \} \sim
{\rm Im}\,( H_i+2A_i^a{\cal G}_a) $, these transformations
map solutions of (\ref{ImHi}) to themselves inside the surface
${\cal G}=0$. One can use $SO(3)$ transformations and diffeomorphisms
to impose the condition $\chi =0$ everywhere, and
$\partial_k\tE^j_a=0$ at a certain point. At this point one must
only check cancellation of second derivatives of $\tE$.
This is straightforward to do by using the equations (\ref{CtoR}),
(\ref{rai}) and the explicit form (\ref{HP5}) of the
constraint ${\cal G}=\Phi^G+i\Phi^L$.

To prove that ${\rm Im}\, H=0$ one can use the Lorentz boosts
to put $\chi =0$. This makes the calculations quite elementary
even without further gauge fixing.

By straightforward calculations one can demonstrate that imaginary
part of the kinetic term $P^j_a\partial_t A_j^a$ is a total
derivative and thus can be discarded in quantization. This is
done in the Appendix C.

As it was advertised at the beginning of this section,
we demonstrated that the complex canonical
variables satisfy the reality conditions on the surface
of the equations (\ref{CtoR}), the second class
constraint (\ref{2class}) and two first class constraints
$\Phi^G$ and $\Phi^L$. Note, that the reality conditions admit
more solutions. For example, one can interchange real and
imaginary parts of $P^j_a$.

\section{Path integral quantization of the Ashtekar gravity}

 In this section we derive a path integral for the Ashtekar gravity
from the one for the Hilbert--Palatini gravity.

 Consider functional (\ref{ZHP}) in an YM-gauge.
\begin{eqnarray}
Z[j,J] =\int {\cal D}\ita {\cal D}\E {\cal D}\omega^a
{\cal D}\chi_a {\cal D}{\cal N}_G {\cal D}{\cal N}_L
{\cal D}\nd^i {\cal D}\tNn {\cal D}c^{\alpha}
{\cal D}\bar c_{\alpha} \delta (f^{\alpha} {+} g^{\alpha})
\nonumber \\
\times  \exp \left( i\int dt\,
(L_{eff}'+j_a^i\eta_i^a+J_i^a\tE^i_a)\right)  \label{ZA1} \end{eqnarray}
We dropp out the sources for the Lagrange multipliers, $\chi$ and
$\omega$. Discussion of the source terms is postponed to the
end of this section.

Since the gauge
fixing functions $g^\alpha$
does not depend on the Gauss and Lorentz Lagrange multipliers,
integration over these Lagrange multipliers gives
$\delta$-functions of the corresponding constraints,
$\delta (\Fg_a )\delta (\Fl_a)$. This means that in fact we are
working on the surface of these constraints.
In
the previous section
it is shown that on this surface
 imaginary part of the Ashtekar action vanishes. Thus one can write
\be L_{eff}'= L_{A}(P,A)-
i\bar c_{\beta}\Bigl( \frac{\partial g^{\beta}}{\partial n^{\alpha}}
\partial_t - \frac{\partial g^{\beta}}{\partial n^{\gamma}}
C^{\gamma}_{\alpha \lambda}n^{\lambda}+\{ \FF_{\alpha} ,f^{\beta}\} \Bigr)
c^{\alpha}
\ee
We assume that complex canonical variables are expressed in terms
of real canonical variables by means of (\ref{CtoR}).

One can integrate over
$\omega^a$ by using the delta function of the Lorentz constraint
$\Phi^L$. This is equivalent in effect to the substitution:
\be \omega^a(\tE,\eta,\chi):=(\delta_{ab}+
\frac{\chi_a\chi_b}{1-\chi^2})\partial_i\tE^i_b+\tE^i_a\eta^b_i\chi_b
-\frac{\chi_a}{1-\chi^2}(\tE^i_b\eta_i^b-\tE^i_b\chi_b\eta_i^c\chi_c)
\plabel{om0}
\ee
The path integral measure is multiplied by
\be \Delta_1={\rm det}^{-1}\X{a}{b}=
\prod_{x,t}\frac1{1-\chi^2}. \plabel{d1} \ee

Now we are ready to change the integration variables in (\ref{ZA1}):
\beq \tE^i_a &\longrightarrow & P^i_a=i\tE^i_a+\eps^{abc}\tE^i_b\chi_c
\nonumber \\
\eta_i^a &\longrightarrow & A_i^a=\xi_i^a-i(\eta_i^a+
\eps^{abc}\xi_i^b\chi_c)   \label{chan}  \eeq
This gives rise to a determinant
\beq \Delta_2&=&{\rm det}^{-1}\left( 1i\delta^i_j\delta_a^b+
\delta^i_j\eps^{abc}\chi_c\right) {\rm det}^{-1}\Bigl( \frac1{2(1-\chi^2)}
\left( -2\delta_i^j\eps^{abc}\chi_c
+\eps^{apq}\tE^j_q\X{p}{b}\Et_i^d\chi_d  \right. \Bigr.  \nonumber \\
&& \left. -\chi_a\eps^{dpq}\Et_i^d\tE^j_q\X{p}{b}\right)
 -i\bigl( \delta_i^j\delta^a_b+\frac1{2(1-\chi^2)}\left(
2\delta_i^j(\delta^a_b\chi^2-\chi_a\chi_b) \right. \bigr.
 \nonumber \\
&& \Bigr. \bigl. \left.
+(1-\chi^2)\tE^j_a\chi_b\Et_i^c\chi_c-\X{a}{b}\Et_i^c\chi_c\tE^j_d\chi_d
\right) \bigr) \Bigr)
=\prod_{x,t} \left( -\frac1{1-\chi^2} \right)   \label{d2}
\eeq

Note, that if all the gauge fixing functions $f$ depend on
the real fields $\chi$ and $\tE$ through $P$ only, the
ghost action becomes degenerate (see (\ref{prop})). This is
a manifestation of the fact that the Lorentz constraint is
"superfluous" in the complex Ashtekar gravity. Therefore,
we must fix corresponding gauge freedom by means of a condition
on $\chi$:
\begin{equation}
\chi^a=\chi^a_{(0)}(\tE ) ,\plabel{chi0}
\end{equation}
where $\chi_{(0)}$ is a given function.

Before integrating over $\chi$ let us rewrite (\ref{chi0}) in a
different form. By inverting the first equation in (\ref{CtoR}),
one obtains
\begin{equation}
\tE^i_a=\left( \frac{\eps^{abc}\chi_c}{1-\chi^2}-i\frac{\delta_{ab}-
\chi_a\chi_b}{1-\chi^2}\right)P^i_b=\pi_a^b(\chi )P^i_b .
\plabel{EP}
\end{equation}
Due to (\ref{chi0}) one can replace $\chi$ by $\chi_{(0)}(\tE )$.
Right hand side of (\ref{EP}) becomes $\tE$ dependent. This
dependence, however, can be removed at least locally be means
of formal power series expansion. As a result, we obtain
\begin{equation}
\tE^i_a=\bar \pi_a^b (P) P^i_b , \plabel{EP1}
\end{equation}
where $\bar \pi$ is a function of $P$ but not of $P^*$, which
depends on a choice of the gauge fixing function $\chi^{(0)}$.
For the present analysis explicit form of $\bar \pi$ is of no
importance. Note, that simple relation $\tE ={\rm Im} P$ would
not work, because it depends both on $P$ and its complex conjugate.

One can replace (\ref{chi0}) by the condition
\begin{equation}
\chi =\chi_{(0)} (\bar \pi P) =\bar \chi (P)\ .
\plabel{chi1}
\end{equation}
The two conditions (\ref{chi0}) and (\ref{chi1}) are equivalent
since they select the same surfaces in the phase space.
However, ghost terms and Jacobian factors appearing due to
delta functions of gauge conditions are different for (\ref{chi0})
and (\ref{chi1}). In the final result these differences
compensate each other, as one can easily show using geometric
interpretation of the Faddeev--Popov determinant.

Let us integrate over $\chi$ with the help of the delta function
$\delta (\chi -\bar \chi (P))$. Since we already changed
variables to $P$ and $A$, no Jacobian factor appears.

Intergation over $P$ and $A$ should be understood as a contour
integration in complex space. One integrates along the lines
defined by the reality conditions and the equations (\ref{chi0})
and (\ref{om0}). As usual, there are real parameters which
label points of the contours in the complex planes. These are
$\tE$ and $\eta$. Since the fields $\omega$ and $\chi$ are already
excluded, we do not integrate over position of the contours.

Consider the ghost action. Integration over $\bar c$ and $c$
gives the following functional determinant:
\begin{equation}
\det
\Bigl( \frac{\partial g^{\beta}}{\partial n^{\alpha}}
\partial_t - \frac{\partial g^{\beta}}{\partial n^{\gamma}}
C^{\gamma}_{\alpha \lambda}n^{\lambda}+\{ \FF_{\alpha} ,f^{\beta}\} \Bigr)
\plabel{cdet}
\end{equation}
Let us separate indices corresponding to the Lorentz boosts:
$\{ \Phi_\alpha \} = \{ \Phi^L_a ; \Phi_\mu \}$,
$\{ f^\alpha \} = \{ \chi^a -\bar \chi^a (P);f^\mu (\chi , P)\}$,
$\{ g^\alpha \} =\{ 0;g^\mu \}$. Greek indices from the middle
of the alphabet correspond to the Gauss law, diffeomorphism and
Hamiltonian constraints. Matrix elements in (\ref{cdet})
contain the following brackets:
\begin{eqnarray}
&&\left\{ \Phi_\mu ,f^\nu (\chi , P)\right\} =
\{ \Phi_\mu ,P\} \frac{\delta f^\nu}{\delta P} +
\frac{\delta \Phi_\mu}{\delta\omega} \frac{\delta f^\nu}{\delta\chi} ,
\nonumber \\
&&\left\{ \Phi_\mu ,\chi -\bar\chi (P)\right\} =
\frac{\delta \Phi_\mu}{\delta \omega} -
\{ \Phi_\mu ,P\} \frac{\delta \bar\chi}{\delta P} ,
\plabel{lines}
\end{eqnarray}
where summation indices are suppressed. Let us multiply the lines
corresponding to $\chi^a -\bar\chi^a$ by $-\delta f^\nu /\delta \chi^a$
and add them to the $f^\nu$ lines. This produces the matrix
elements:
\begin{eqnarray}
&&\frac{\partial g^{\nu}}{\partial n^{\mu}}\partial_t -
\frac{\partial g^{\nu}}{\partial n^{\rho}}
C^{\rho}_{\mu \sigma}n^{\sigma}+
\{ \Phi_\mu ,P\} \left(  \frac{\delta f^\nu}{\delta P}
+\frac{\delta f^\nu}{\delta \chi} \frac{\delta \bar\chi}{\delta P}
\right) = \nonumber \\
&&\frac{\partial g^{\nu}}{\partial n^{\mu}}\partial_t -
\frac{\partial g^{\nu}}{\partial n^{\rho}}
C^{\rho}_{\mu \sigma}n^{\sigma}+
\left\{ \Phi_\mu^{[C]}, f^\nu (\bar\chi (P),P) \right\}_C .
\plabel{fchiP}
\end{eqnarray}
$\Phi_\mu^{[C]}$ is the Ashtekar constraint corresponding to $\Phi_\mu$,
${\rm Re}\, \Phi_\mu^{[C]}=\Phi_\mu$. In the last line we used that
$\{ \Phi_\mu ,P\} = \{ \Phi_\mu^{[C]} ,P\}_C$ due to (\ref{prop}).
The equation (\ref{fchiP}) means
that one replace $\chi$ by $\bar\chi$ in the gauge fixing functions
$f^\nu$.

Consider the two columns in (\ref{cdet}) corresponding to the
Gauss law and Lorentz constraints. Due to (\ref{prop})
$\{ \Phi^G ,f(P)\} =i \{ \Phi^L,f(P)\}$. Therefore, by multiplying
the column with $\Phi^G$ by $-i$ and adding it to the column with
$\Phi^G$ one obtains zeros almost everywhere, except for the lines
corresponding to the gauge conditions $\chi^a -\bar\chi^a (P)$.
As a result, one can represent the determinant (\ref{cdet}) as
a product of two determinants:
\begin{equation}
\Delta_3 \ \det \left(
\frac{\partial g^{\nu}}{\partial n^{\mu}}\partial_t -
\frac{\partial g^{\nu}}{\partial n^{\rho}}
C^{\rho}_{\mu \sigma}n^{\sigma}+
\left\{ \Phi_\mu^{[C]}, f^\nu (\bar\chi (P),P) \right\}_C \right) ,
\plabel{newdet}
\end{equation}
where
\be \Delta_3= \det  \{ \Phi_a^L-i\Phi_a^G ,\chi^b\} =
{\rm det}\left( \X{a}{b}+i\eps^{abc}\chi_c\right)=
\prod_{x,t}(1-\chi^2)^2 \label{d3} \ee

From the expressions (\ref{d1}), (\ref{d2}) and (\ref{d3}) one can
see that all $\Delta$'s cancel each other up to an overall minus
sign which can be absorbed in reversed orientation of the contour
of the $A$-integration. The path integral is now rewritten in terms
of the Ashtekar variables:
 \be Z[\bar j,\bar J] =\int_R {\cal D}\A {\cal D}P^i_a
 {\cal D}\nd^i {\cal D}\tNn {\cal D} A_0^a {\cal D}c^{\mu}
{\cal D}\bar c_{\mu} \delta (f^{\mu} {+} g^{\mu})
e^{i\int dt\,
(L_{eff}'+\bar j_a^iA_i^a+\bar J_i^aP^i_a)}  \label{ZA2} \ee
where
\be L_{eff}'= L_{A}-
i\bar c_{\nu}\Bigl(
\frac{\partial g^{\nu}}{\partial n^{\mu}}\partial_t -
\frac{\partial g^{\nu}}{\partial n^{\rho}}
C^{\rho}_{\mu \sigma}n^{\sigma}+
\left\{ \Phi_\mu^{[C]}, f^\nu (\bar\chi (P),P) \right\}_C
\Bigr) c^{\mu}
\ee
The subscript $R$ means contour integration in complex spaces along lines
defined by the reality conditions. Integration over ${\cal N}_L$ (which
is essentially an imaginary part of $A_0$) has been already performed
to produce a delta function of the Lorentz constraint. This delta
function, in turn, has been used to integrate over $\omega$. Thus
in (\ref{ZA2}) we integrate over real part of $A_0$. This integral
gives $\delta (\Phi^G )=\delta ({\cal G})$. The equation
${\cal G}=0$ can be considered as a complex equation because
${\rm Im}\, {\cal G}=0$ is supplied by the reality conditions.
The same is true for the gauge conditions $f^{\mu} {+} g^{\mu}=0$.
A fascinating property of these complex delta functions is
possibility to integrate over complex variables without explicit
transition to real coordinates on a contour.

By comparing (\ref{algA}) and (\ref{alg}), one can see that
$C^{\rho}_{\mu \sigma}$ are just structure constants of the
Ashtekar gravity (Note, that this property does not hold in
the variables used by Henneaux \cite{He})
Therefore, the ghost term in (\ref{ZA2}) produces
the ordinary Faddeev--Popov determinant for the Ashtekar gravity.
The path integral (\ref{ZA2}) coincides with what one would write naively
just ignoring any Jacobian factors which may arise from the reality
conditions and fixing the Lorentz gauge freedom.
Some remarks are in order. First of all, the result
(\ref{ZA2}) is valid for a certain class of gauges only. We are not
allowed to impose gauge condition on $A_0^a$. This restriction is
needed (i) to cancel contributions to the path integral
of the second order structure functions (which are zero for the
Ashtekar gravity \cite{AshMaTo}),
and (ii) to ensure delta functions
of the complex Gauss law constraint. While (i) seems to depend on
a particular choice of basic variables and constraints because rank of
and algebra is not an invariant, the second point (ii) looks more
fundamental. The complex Gauss law constraint is needed to prove
vanishing of imaginary part of the Ashtekar action. We are not
allowed to impose gauge conditions on the connection variables.
The ultimate reason for this is that the last line of (\ref{prop})
is not true if we replace $P$ by $A$. This restriction will receive
a natural explanation in the next section in a framework of the
Faddeev path integral. In all other respects the gauge
conditions $f^\alpha +g^\alpha$ are arbitrary. For a given set of
admissible YM gauges one can first express $\chi^a$ from three
of them and then denote the remaining gauge conditions by
$f^\mu +g^\mu$.

Path integral for the Ashtekar gravity was previously considered
by the present authors and I.~Grigentch in the one--loop
approximation over de Sitter background \cite{GV} and for
the Bianchi IX finite dimensional model \cite{AGV}. In these
simple cases the reality conditions do not lead to any
Jacobian factors if one uses gauge conditions of the YM type.
We observed also that one runs into troubles if gauge conditions
are imposed on the connection variables.

Using of this or that gauge condition is just a matter of
convenience. In principle, it is enough to formulate the
path integral in just one gauge. All physical results are to be
gauge independent. However, extension of our results for arbitrary
gauge conditions still poses an interesting problem from both
technical and aesthetic points of view.

Note, that we excluded sources for $\chi$, $\omega$ and
Lagrange multipliers. Sources for $\chi$ and $\omega$ are
not needed because in the present formulation these fields
are absent. Moreover, $\chi$ and $\omega$ can be considered
as composite fields.
Sources for $\tN$ and $\nd$ can be easily restored without
any modification in our procedure. Therefore, we have enough
sources to describe any Green functions of the four-metrics
and three-dimensional connections. If, however, we introduce
a source for $A_0^a$, it penetrates into the delta functions
of the Gauss law and Lorentz constraints and destroys reality
of the Ashtekar action. Green functions of $A_0$ are not defined
in our approach. At the last step we introduced sources $\bar J$
and $\bar j$ for $P$ and $A$. This makes exponential in (\ref{ZA2})
complex. Thus, strictly speaking, the path integral is not well
defined, even though all finite order Green functions do exist.
If one wishes to be on a safe side, one can easily
return to the original sources $J$ and $j$ for $\tE$ and $\eta$.

\section{The Faddeev path integral}
In this section we give a more simple derivation of the Faddeev path
integral \cite{Faddeev} for the Ashtekar gravity,
which does not rely upon heavy machinery of the BRST quantization.
This also seems to be a proper place to discuss triad form of
the reality conditions.
For a dynamical system with canonical variables $q^s,p_s$,
first class constraints $\Phi_a$ and weakly vanishing Hamiltonian,
such as the Hilbert--Palatini gravity,
the Faddeev path integral reads:
\begin{equation}
Z=\int {\cal D}q {\cal D}p {\cal D}n F \delta (f^{\alpha})
\exp \left( i\int dt\, (\dot{q}^sp_s +n^\alpha\Phi_\alpha )\right)
\plabel{ZFP}
\end{equation}
where $f^\alpha$ are gauge fixing functions of the dynamical
variables. $F$ is the Faddeev--Popov determinant,
$F=\det \{ \Phi_\alpha ,f^\beta \}$. We do not show the source terms
explicitly. The expression (\ref{ZFP}) can be obtained by from
the path integral (\ref{ZHP}) by choosing $g^\alpha =0$ and
integrating over the ghost fields $c$ and $\bar c$. Of course,
starting point of the original derivation \cite{Faddeev} of the
Faddeev path integral was not the BRST approach.

To make the presentation as simple as possible, we fix
Lorentz boosts by the condition
\begin{equation}
\chi =0 . \plabel{chic}
\end{equation}
Now we integrate over ${\cal N}_L^a$, $\chi$ and $\omega$. Again,
integration over $\omega$ is equivalent to the following
substitution:
\begin{equation}
\omega_a :=\partial_j \tE^j_a .\plabel{sub-om}
\end{equation}
If the remaining gauge fixing conditions $f^\mu$ are functions of
$\tE$ only, the Poisson brackets $\{ f^\mu ,\Phi_L^a\}$ vanish
on the surface (\ref{chic}).
Hence the Faddeev--Popov determinant takes the form
\begin{equation}
F=\det \, \{ f^\mu (\tE ),\Phi_\nu \} =
\det\, \{ f^\mu (-iP), \Phi^{[C]}_\nu \}_C
\plabel{FP88}
\end{equation}

The gauge (\ref{chic}) means that we are using
reality conditions in the triad form
\begin{equation}
{\rm Re}\, P_a^i=0, \qquad {\rm Re}\,(\partial_t P_a^i)=0
\plabel{trirc}
\end{equation}
instead of the metric reality conditions (\ref{1rc}) and
(\ref{2rc}).

The change of variables $(\tE ,\eta )\to (P,A)$ gives unit
Jacobian factor.
Our prove of vanishing of imaginary part of the Ashtekar
action is still valid. Hence we arrive at the path integral
for the Ashtekar gravity in the Faddeev form:
\begin{equation}
Z=\int_R {\cal D}P {\cal D}A {\cal D}\tN {\cal D}\nd {\cal D}A_0
 F \delta (f^{\mu}(-iP))
\exp \left( iS_{A} \right)
\plabel{FPZ88}
\end{equation}
where subscript $R$ means now that the contour of integration is
defined by the reality conditions (\ref{trirc}).
Of course, most of the comments of the previous section
apply here also.

\section{Discussion}
Main result of the present paper is the path integral
(\ref{ZA2}) for the Ashtekar gravity, which is a kind
of contour integral. As a byproduct, we also constructed
the BRST quantization of the Hilbert--Palatini gravity.
Main features of our approach were discussed in detail
in the section VI. Here we speculate on perspectives
of this approach.

The path integral (\ref{ZA2}) is obtained
with certain restrictions on possible gauge conditions.
In principle, one can transform (\ref{ZA2}) to any other
gauge by means of the Faddeev--Popov trick \cite{FaPo}.
However, this trick is not so easy to implement in the
present context due to reality conditions and quite
unusual rules of the functional integration. Perhaps
restrictions on the gauge conditions may be weakened
or even lifted altogether. Anyhow, one should formulate
criteria of admissibility of gauge conditions for the
Ashtekar gravity in terms of the Ashtekar variables without
referring to the Hilbert--Palatini gravity. This definitely
will not be easy to do. In general, a function of $P$
is complex valued. Therefore, a condition $f=0$ implies
two real gauge fixing conditions ${\rm Re} f=0$ and
${\rm Im}f=0$ even if reality conditions are taken into
account. Even the requirement that a given set of gauge
conditions removes correct number of degrees of freedom
looks quite non-trivial.
One may hope to overcome these difficulties
by using the generalized Wick rotation \cite{Wick}.

We must admit that for degenerate triad our analysis is
incomplete. This reflects a well known problem of the
Ashtekar gravity which exists already at the classical
level.

An intriguing feature of (\ref{ZA2}) is that it is a contour
integral. The contour of integration can be deformed as far
as the reality conditions allow (This corresponds
to arbitrariness of gauge fixing in the Hilbert--Palatini
action.) One may hope, that certain
deformations are possible even beyond these limits.
If this is really so, some interesting properties of
quantum gravity can manifest themselves.

\section*{Acknowledgments}

This work was supported by the Russian Foundation
for Fundamental Research, grant 97-01-01186, and
by GRACENAS through grant 97-0-14.1-61 (D.V.) and Young
Investigator Program (S.A.).

\appendix
\section{}

Let us solve the second class constraint (\ref{2class}).
The matrix ${\cal A}_{AB}$ is defined by the $r^2$ terms in the
Hamiltonian constraint $\Phi^H$. We have:
\begin{equation}
\lambda^A {\cal A}_{AB}\lambda^B =r_b^c
{\cal A}_{bb'}^{cc'} r_{b'}^{c'}
\plabel{A}
\end{equation}
where $r_b^c=r_j^c\tE^j_b$.
We can identify non-dynamical components of the connection
$\lambda_A$ with the symmetric matrices $r_a^b$.
The operator
\begin{equation}
{\cal A}_{bb'}^{cc'}=(1-\chi^2)\eps^{abb'}\eps^{dcc'}X_{ad},
\quad X_{ad}=\delta_{ad}-\chi_a\chi_d
\plabel{A2}
\end{equation}
acts on the space of symmetric $3\times 3$ matrices. One
can represent it in the following form:
\begin{equation}
{\cal A}_{bb'}^{cc'}=(1-\chi^2)^2 (X^{bc}X^{b'c'}-X^{bc'}X^{cb'})
\plabel{A3}
\end{equation}
where $X^{bc}$ is inverse of $X_{bc}$. Inverse of (\ref{A3})
is easily found to be
\begin{equation}
({\cal A}^{-1})_{bb'}^{cc'}= (1-\chi^2)^{-2}(\frac 12
X_{bc}X_{b'c'}-X_{bc'}X_{cb'}) \plabel{A-1}
\end{equation}
Linear part of the Hamiltonian constraint reads:
\begin{equation}
{\cal B}_A\lambda^A=\eps^{abc}\tE^i_b(\tE^j_cr_d^a
(\partial_i\Et_j^d)(1-\chi^2)+r_c^d\chi_d\partial_i\chi_a
-(1-\chi^2)\chi_ar_c^d\eta_i^d+X_{ag}\eta_i^gr_c^d\chi_d)
={\cal B}_a^br_b^a \label{B}
\end{equation}
Note, that (\ref{B}) does not contain derivatives of $r_a^b$.
Hence the
second class constraint (\ref{2class})
can be solved for $r_a^b $:
\begin{equation}
r_d^c =\frac 12 \left[ ({\cal A}^{-1})^{ca}_{db}+
({\cal A}^{-1})^{cb}_{da} \right]
{\cal B}_a^b
\plabel{sol2}
\end{equation}
Substitution of (\ref{A-1}) and(\ref{B}) in (\ref{sol2})
gives the expression
(\ref{rai}). The Hamiltonian constraint takes the form
$\Phi^H=\Phi^H_0 +{\cal B}_a^b({\cal A}^{-1})_{bd}^{ac}{\cal B}_c^d$,
which is written explicitly in
(\ref{newH}).
\section{}

In this Appendix we define structure functions
$U^{(n)}$ of the Hilbert--Palatini gravity.
For $n=0$ and $n=1$ they are
\begin{equation}
U^{(0)}_\alpha =\FF_\alpha ,\quad
U^{(1)\gamma}_{\alpha\beta}=
-\frac 12 C^{\gamma}_{\alpha\beta} , \plabel{U01}
\end{equation}
with $C^{\gamma}_{\alpha\beta}$ defined by the algebra (\ref{alg})
through the relation $\{ \FF_\alpha ,\FF_\beta \} =
C^{\gamma}_{\alpha\beta}\FF_\gamma$. Higher order structure functions
are defined through repeated Poisson brackets of the constraints
\begin{equation}
2U^{(2)\xi\eta}_{\alpha\beta\gamma}\Phi_\eta =
D^{(1)\xi}_{\alpha\beta\gamma}=
\frac 12 \left( \{ \FF_{\alpha},C^\xi_{\beta\gamma } \} -
C^\delta_{\beta\gamma}C^\xi_{\alpha \delta} \right)_{[\alpha \beta \gamma ]}
\plabel{U2}
\end{equation}
where $[{\alpha_1\cdots \alpha_n}]$ means antisymmetrization in
${\alpha_1\cdots \alpha_n}$ with the weight $1/n!$. In actual
calculations it is convenient to replace antisymmetrization
by multiplication by anticommuting ghosts.
The indices $\alpha ,\beta ,\dots$ denote constraints at different
coordinate points. Therefore, antisymmetrisation over coinciding
indices does not necessarily give zero.

If less than two indices among $\alpha$, $\beta$ and $\gamma$
correspond to the Hamiltonian constraint, the structure
functions $C$ in (\ref{U2}) become field independent structure
constants, and the second order structure functions
$U^{(2)\xi\eta}_{\alpha\beta\gamma}$ vanish
by virtue of ordinary Bianchi identities . Hence, one must
calculate only the structure functions with a pair of
indices, say $\beta$ and $\gamma$, corresponding to the
Hamiltonian constraint. From now on, an index representing the
Hamiltonian constraint will be denoted by $0$,
$\Phi^H\equiv\Phi_0$. We put $\gamma =\beta =0$.

It is convenient to introduce a connection field of the Lorentz
group $SO(3,1)$: $A^p_i=(\ksi,\eta_i^x+\eps^{xgf}\xi_i^g\chi_f)$,
$p=1,...,6$.
$f^r_{pq}$ will denote structure constants of corresponding Lie
algebra.

From (\ref{alg}) it is clear that the canonical momenta enter
the first order structure functions $C$ through the vector
$K^j[n,m]=(n\partial_i m -\partial_i n \, m)K^{ij}$, where $K^{ij}$
is defined in (\ref{Kij}). Later $n$ and $m$ will be replaced by
ghost fields. Thus an order is essential. $n$ always precedes $m$.
The tensor $K$ has the following Poisson brackets with the
constraints:
\begin{eqnarray}
&&\{ \Fg_a ,K^{ij}\}=\{ \Fl_a ,K^{ij}\} =0,\quad
\{ c^0\FF_0, K^j[c^0,c^0] \}=0 , \nonumber \\
&& \{ c^k\Fd_k, K^{ij}\} =2(2K^{ij}\partial_k c^k +c^k\partial_k K^{ij}-
\partial_k c^j K^{ik}-\partial_k c^i K^{kj} )
\plabel{comK} \end{eqnarray}
where contraction with anticommuting ghosts $c$ is used for
antisymmetrization  in corresponding indices.

Let us calculate $c^\alpha D^{(1)\xi}_{\alpha 00'}c^0(x)c^0(x')$.
Consider various cases for $\alpha$. If $\Phi_\alpha =
\Phi_p=(\Phi^G,\Phi^L)$ and $\Phi_\xi =\Phi_0(=\Phi^H)$ or
$\Phi_\xi =\Phi^D$ this quantity vanishes due to (\ref{comK}).
For $\Phi_\xi=\Phi_q$ one obtains
\begin{equation}
c^p D^{(1)q}_{p00'}c^0(x)c^0(x')=
\frac 23 K^j[c_0(x),c_0(x')]
\left( -\{ c^p \FF_p,A^q_j \} +f^q_{rp}A_j^rc^p+
\partial_jc^q\right) \delta (x,x')
\plabel{Dp}
\end{equation}
As a part of our summation convention we assume integration
over all continuous coordinates here and in the equations
bellow.
The expression (\ref{Dp}) is zero due to (\ref{comm}).
This implies that $U^{(2)\xi \eta}_{00p}=0$.

Let us put $\Phi_\alpha =\Phi^D_i$. We are to evaluate:
\begin{equation}
c^i D^{(1)\xi}_{i0'0''}c^0(x')c^0(x'')=
\frac 16 c^i\left( \{ \FF_{i},C^\xi_{0'0''} \} -
C^\beta_{0'0''}C^\xi_{i\beta }
-2C^\beta_{i0'}C^{\xi}_{0''\beta} \right)c^0(x')c^0(x'') .
\plabel{Di}
\end{equation}
First we observe that the only non-vanishing function $C$ with
zero upper index is $C^0_{oi}$. This immediately gives
vanishing of (\ref{Di}) for $\xi =0$. Other components of
(\ref{Di}) vanish due to (\ref{comm}) and (\ref{comK}).

For $\alpha =0$ we have:
\begin{eqnarray}
&&c^0(x)D^{(1)0}_{00'0''}c^0(x')c^0(x'')=
-\frac 12 c^0(x)C^i_{00'}C^{0}_{i0''}c^0(x')c^0(x'')
=8c^0\partial_ic^0\partial_jc^0K^{ij}\delta (x,x')\delta (x',x'')=0
\nonumber \\
&&c^0(x)D^{(1)i}_{00'0''}c^0(x')c^0(x'')=
\frac 12 c^0(x) \{ \FF_{0},C^i_{00 } \}
c^0(x')c^0(x'')=0  \nonumber \\
&&c^0(x)D^{(1)p}_{00'0''}c^0(x')c^0(x'')=
-2 \{ c^0(x) \Phi^H (x), K^j [c^0(x'),c^0(x'')]A_0^p\}
\end{eqnarray}
where the first line is zero due to contraction of a symmetric tensor
with an antisymmetric one. In the second line we used second of the
equations (\ref{comK}).

To calculate the remaining components of $D^{(1)}$ the following
brackets are needed:
\beq
\left\{ c^0 \FF_0,K^j[c^0,c^0]\xi_j^a\right\}
&=&2c^0\partial_ic^0
\partial_kc^0\left( (\tE^i_a\tE^k_b-\tE^k_a\tE^i_b)\Fg_b \right.
\nonumber \\
 && \left. +(\tE^i_a\tE^k_g-\tE^k_a\tE^i_g)\eps^{gfb}\chi_f\Fl_b \right)
 \nonumber \\
  \left\{ c^0 \FF_0,K^j[c^0,c^0](\eta_j^a+\eps^{abc}\xi_j^b\chi_c)
  \right\} &=&2c^0\partial_ic^0 \partial_kc^0\left(
  -\eps^{ade}\chi_d(\tE^i_e\tE^k_b-\tE^k_e\tE^i_b)\Fg_b \right.
\nonumber \\
 && \left. -\eps^{ade}\chi_d
 (\tE^i_e\tE^k_g-\tE^k_e\tE^i_g)\eps^{gfb}\chi_f\Fl_b \right)
 \eeq

By introducing a $3\times 6$ matrix field $\tE^i_p=(\tE^i_a,\eps_{abc}
\tE^i_b\chi_c)$, $p=1,...,6$, one can represent the non-vanishing
second order structure functions in an elegant form:
\begin{equation}
c^0(x) U^{(2)pq}_{00'0''}c^0(x')c^0(x'')=
-8c^0\partial_ic^0\partial_kc^0
   \tE^i_p\tE^k_q \delta (x,x')\delta (x',x'')
\plabel{sf2} \end{equation}

Third order structure functions are defined as:
  \be 3U^{(3)\xi \eta \lambda }_{\alpha \beta \gamma \delta }\FF_{\lambda}
 =\left( -
   \{ U^{(2)\xi \eta }_{\alpha \beta \gamma }, \FF_{\delta} \}
   -\frac18 \{ C^{\xi}_{\alpha \beta }, C^{\eta}_{\gamma \delta } \}
   +\frac32C^{\lambda}_{\alpha \beta}
   U^{(2)\xi \eta }_{\gamma \delta \lambda }
   +2U^{(2)\xi \lambda }_{\alpha \beta \gamma }C^{\eta}_{\delta \lambda }
   \right)_{[\alpha \beta \gamma \delta]}^{[\xi \eta]}
\plabel{U3}
\ee

  As before, only the functions with $\alpha =\beta =\gamma =0$
could be non-zero. By straightforward calculations one can
demonstrate that they vanish as well. There are no non-zero third or
higher order structure functions in the Hilbert--Palatini gravity.

\section{}
In this Appendix we prove that imaginary part of the kinetic term
of the Ashtekar action
 (\ref{HP3}) vanishes for the fields (\ref{CtoR}) provided the real
canonical variables satisfy the
second class constraints (\ref{2class}) and
the Gauss and Lorentz constraints.

Consider the kinetic term:
\beq {\rm Im}\, A_i^a\partial_t P^i_a&=&\left( \left( \delta_{ab}(1-\chi^2)+
\chi_a\chi_b\right) \xi_i^b-\eps^{abc}\chi_b\eta_i^c\right)
\partial_t \tE^i_a \nonumber \\
&& -\left( \eps^{abc}\eta_i^b\tE^i_c+\chi_a\xi_i^b\tE^i_b-
\xi_i^a\tE^i_b\chi_b\right) \partial_t \chi_a,\plabel{Imk}
\eeq
where the expressions (\ref{CtoR}) were substituted. By making
use of the constraints one can rewrite (\ref{Imk}) in the
following form.
\beq {\rm Im}\, A_i^a\partial_t P^i_a&=&-\frac12 \partial_t\tE^i_a\left[
 \eps^{abc}\tE_b^k \tE^j_c \X{g}{f} \Et_i^g \partial_k \Et_j^f \right.
\nonumber \\
&&- \Et_i^a \eps^{dbc} \tE^k_b \tE^j_c \X{d}{f} \partial_k\Et_j^f
+\eps^{dbc} \tE_b^k\tE_c^j \X{d}{g} \Et_i^g \partial_k\Et_j^a
\nonumber \\
&&\left. -\eps^{abc} \partial_j \tE^j_c\X{b}{g}  \Et_i^g
-\Et_i^a\eps^{dbc}\Et^j_b\partial_j\chi_d\chi_c
+2\chi_g\Et_i^g\eps^{abc} \tE_c^j\partial_j\chi_b \right] \nonumber \\
&&-\frac12 \partial_t\chi_a \left[ \eps^{abc}\tE^k_b\tE^j_c\chi_g
\partial_k \Et_j^g +\eps^{dbc}\tE^k_b\tE^j_c\chi_d\partial_k\Et_j^a+
\eps^{abc}\partial_j\Et^j_b\chi_c +2\eps^{abc}\tE^j_b\partial_j\chi_c
\right]  \nonumber \\
&=&\frac12 \partial_t\left( \eps^{dbc}\tE^k_b\tE^j_c\X{d}{g}\right)
\partial_k\Et_j^g -\frac12\partial_t\Et_j^g \partial_k\left( \eps^{dbc}
\tE^k_b\tE^j_c\X{d}{g}\right) \nonumber \\
&& +\frac12\partial_t(\tE_a^i\chi_b)\eps^{abc}\partial_i\chi_c -
\frac12\eps^{abc}\partial_t\chi_c\partial_i(\tE^i_a\chi_b) \nonumber \\
&=&\frac12 \partial_t\left( \eps^{dbc}\tE^k_b\tE^j_c\X{d}{g}
\partial_k\Et_j^g \right)
-\frac12\partial_k \left( \eps^{dbc}\tE^k_b\tE^j_c\X{d}{g}
\partial_t\Et_j^g\right) \nonumber \\
&& +\frac12\partial_t\left(\eps^{abc}\tE_a^i\chi_b\partial_i\chi_c\right)-
\frac12\partial_i\left( \eps^{abc}\tE^i_a\chi_b\partial_t\chi_c \right)
\nonumber \eeq
Thus the imaginary part of the kinetic term is a
total derivative and can be neglected.


\begin{references}
\bibitem{Ash-1}
A.~Ashtekar, Phys. Rev. Lett. {\bf 57}, 2244 (1986);
Phys. Rev. {\bf D 36}, 1587 (1987).
\bibitem{MeNi}
S.~Melosh, H.~Nicolai, Phys. Lett. {\bf B416}, 91 (1998).
\bibitem{Rov}
C.~Rovelli, {\it Loop Quantum Gravity}, electronic journal
Living Reviews, gr-qc/9710008.
\bibitem{Abook}
A.~Ashtekar,
{\it Lectures on non-perturbative canonical gravity. Notes
prepared in collaboration with R.~Tate}
(World Scientific, Singapore, 1991).
\bibitem{Peldan}
P.~Peldan, Class. Quantum Grav. {\bf 11}, 1087 (1994).
\bibitem{AshMaTo}
A.~Ashtekar, P.~Mazur and C.G.~Torre, Phys. Rev. {\bf D36},
2955 (1987).
\bibitem{FaPo}
L.D.~Faddeev and V.N.~Popov, Phys. Lett. {\bf B25}, 29 (1967).
\bibitem{He}
M.~Henneaux, Phys. Rev. {\bf D24}, 986 (1983).
\bibitem{NeTe}
J.E.~Nelson and C.~Teitelboim, Ann. Phys. (NY) {\bf 116}, 86
(1978).
\bibitem{ABaJo}
A.~Ashtekar, A.P.~Balachandran and S.G.~Jo,
Int. J. Mod. Phys. {\bf A4}, 1493 (1989).

\bibitem{Sam}
J.~Samuel, Pramana--J. Phys. {\bf 28}, L429 (1987);\\
T.~Jacobson and L.~Smolin, Class. Quantum Grav. {\bf 5}, 583 (1988).
\bibitem{Ben}
I.~Bengtsson, Int. J. Mod. Phys. {\bf A4}, 5527 (1989).
\bibitem{Wal}
R.P.~Wallner, Phys. Rev. {\bf D46}, 4263 (1992).
\bibitem{ARoTa}
A.~Ashtekar, J.D.~Romano and R.S.~Tate, Phys. Rev. {\bf D40},
2572 (1989).
\bibitem{Imm}
G.~Immirzi, Class. Quantum Grav. {\bf 10}, 2347 (1993).
\bibitem{YoSh}
G.~Yoneda and H.~Shinkai, Class. Quantum Grav. {\bf 13}, 783 (1996).
\bibitem{Dirac}
P.A.M.~Dirac, {\it Lectures on quantum mechanics} (Yeshiva University,
NY, 1964).
\bibitem{Kh}
V.~Khatsymovsky, Phys. Lett. {\bf B 394}, 57 (1997).

\bibitem{BFV}
E.S.~Fradkin and G.A.~Vilkovisky, Phys. Lett.
                 {\bf B55}, 244 (1975);\\
I.A.~Batalin and G.A.~Vilkovisky, Phys. Lett. {\bf B69}, 309
                 (1977);\\
E.S.~Fradkin and T.E.~Fradkina, Phys. Lett. {\bf B72}, 343 (1978).
\bibitem{Henneaux}
M.~Henneaux, Phys. Rep. {\bf 126}, 1 (1985).
\bibitem{GV}
I. Grigentch and D. Vassilevich, Int. J. Mod. Phys. {\bf D4},
581 (1995).
\bibitem{AGV}
S. Alexandrov, I. Grigentch and D. Vassilevich, Class. Quantum Grav.
{\bf 15}, 573 (1998).
\bibitem{Faddeev}
L.D.~Faddeev, Teor. Mat. Fiz. {\bf 1}, 3 (1969).
\bibitem{Wick}
T.~Thiemann, Class. Quantum Grav. {\bf 13}, 1383 (1996), \\
A.~Ashtekar, Phys. Rev. {\bf D53}, 2865 (1996).

\end{references}
\end{document}